# Ionic Transport and Selectivity of Electrokinetically-Actuated Non-Newtonian Flows within a pH-Regulated Rectangular Nanochannel

Mohammad Ali Vakili , Morteza Sadeghi, Mohammad Hassan Saidi, Ali Moosavi

*Center of Excellence in Energy Conversion (CEEC), School of Mechanical Engineering, Sharif University of Technology, Tehran 11155-9567, Iran*

Corresponding Author: M. H. Saidi (saman@sharif.edu)

# Ionic Transport and Selectivity of Electrokinetically-Actuated Non-Newtonian Flows within a pH-Regulated Rectangular Nanochannel

**ABSTRACT**. In the present study, the ionic transport and selectivity of electrokinetically-driven flow of power-law fluids in a long pH-regulated rectangular nanochannel are analyzed. The electrical potential and momentum equations are numerically solved through a finite difference procedure for a non-uniform grid. Non-linear Poisson-Boltzmann equation along with the association/dissociation reactions on the surface is considered. In addition, numerical simulations with the finite element method in 3D space are performed to compare the results with those obtained from 2D analysis. Moreover, an analytical solution under Debye-Hückel approximation for the limiting case of a slit nanochannel is derived and its results are compared with those obtained from numerical simulations. It is shown that the channel aspect ratio can influence all the physicochemical parameters. It is observed that the mean velocity and the convective ionic conductance are strong descending functions of the flow behavior index. By investigating the non-Newtonian fluid behavior effect, it is revealed that its impact on the ionic conductance becomes significant at high values of the solution pH and its variation can alter the anionic transport direction inside the nanochannel. Moreover, it is shown the flow behavior index can strongly influence the ion selectivity of the nanochannel and its variation can be used to let the selectivity go through its maximum as a function of pH.

## 1. Introduction

The transport of ions within a nanopore or nanochannel has received remarkable attention due to its versatile applications such as bio-nanoparticles transport and sensing [1-8], sensing/characterizing of analytes [9-12], ion rectification [13-18], ionic gates [19,20], energy conversion [21-24], and sea water desalination [25,26]. It has been demonstrated that the surface charge properties of the nanochannel wall have strong influence on the ionic conductance [27,28] and depend on the solution properties (such as pH and salt concentration) in nanochannels made of functional materials such as glass and silicon [29,30]. Therefore, the ion transport in nanofluidics can be regulated by adjusting the solution pH and consequently modulating the surface charge properties at the solid/liquid interface of these nanofluidic devices [31,32].

Flow movement of an electrolyte solution relative to a charged surface in response to an externally applied electric field is referred to Electroosmosis. The fluid mobilization in electroosmotic flow (EOF) arises from a redistribution of free ions in the solution because of surface charges, causing the formation of an electric double layer (EDL) adjacent to the surface [33-36]. In attempt to investigate the electrokinetic transport in charge regulated media, Hsu et al. [37] derived approximate analytical expressions for the electroosmotic flow in a slit microchannel. They showed that the rate and the direction of the flow can be controlled by the solution pH and the density ratio of acidic and basic functional groups of the surface. Liu et al. [38] obtained analytical solution for electroosmotic flow in a charged regulated circular channel assuming Debye-Hückel approximation and linearized form of the surface charge density expression. Ma et al. [39] considered electroosmotic flow in a pH-regulated cylindrical nanopore and derived analytical expressions for the ionic conductance. The mentioned study was extended by Huang et al. [40] to account for the effects of EDL overlap and Stern layer effects. They showed that the effects of the Stern layer on the ionic conductance are significant at high values of the solution pH. In another work, Ma et al. [41] analyzed electroosmotic flow in a nano slit considering the EDL overlap and Stern layer effects. They demonstrated that the EDL overlap effect is significant at small nanochannel height, low salt concentration, and medium low values of the solution pH. Tseng et al. [42] and Mei et al. [43] examined the effects of the temperature and the buffer solutions on the elecroosmotic flow in pH-regulated nanochannels, respectively. Alizadeh et al. [44] investigated the electrokinetic transport in a pH-regulated nano slit considering a modified electrical triple layer (ETL) model. Ma et al. [45] developed simplified analytical truncated model for estimating the nanochannel conductance by neglecting the contribution of coions to the potential field. They showed that the new model is suitable for long nanochannels and symmetric electrolytes under the condition of EDL overlap. In two more recent researches, Sadeghi et al. [46,47] investigated the effect of the geometry on the ionic conductance and the electroosmotic flow in nanochannels assuming Debye-Hückel approximation and linearized form of the surface charge density expression. They showed that the geometrical configuration has a significant role in determination of the ionic conductance, the surface charge density, the electrical potential and the velocity fields.

Since many biofluids encountered in chemical and biomedical applications show a non-linear behavior that is distinctly different from the Newtonian fluids [48,49], investigating the behavior of such fluids under the influence of the electroosmotic force is of practical importance for accurate design and operation of various micro/nanofluidic devices. Modeling of electrokinetically-driven flows of non-Newtonian fluids in microchannels has been studied extensively. Chakraborty [50] examined electroosmotic flow of power-law fluids in microchannels by means of a semi-analytical mathematical model. Zhao and coworkers [51,52] obtained expressions for the Helmholtz-Smoluchowski electroosmotic velocity of power-law fluids at small and high zeta potentials. Vasu and De [53] and Choi et al. [54] investigated the electroosmotic flow of power-law fluids in a slit microchannel. Babaie et al. [55] numerically analyzed the electroosmotic flow of power-law fluids in slit microchannels in the presence of pressure gradient. Using the same rheological model, Shamshiri et al. [56] and Cho et al. [57] investigated the electrokinetic flow of non-Newtonian fluids in annulus and wavy microchannels, respectively. Vakili et al. [58,59] examined the electroosmotic flow of power-law fluids in rectangular microchannels. Yazdi et al. [60] performed a numerical investigation to analyze the ionic size effects on electroosmotic flow of power-law fluids in rectangular microchannels. The viscoelastic constitutive equations have also received much attention in study of electrokinetic flows in microchannels [61-67]. The available literature indicates that the existing research related to modeling of non-Newtonian fluids in nanofluidics is rare. Matin et al. [68] investigated the effect of biofluids rheological behavior on electroosmotic flow and ionic current rectification in conical nanopores. In this research, the biofluid is assumed to behave as a power-law fluid and Nernst−Plank equation for ionic concentration is considered. Jafari et al. [69] analyzed the electroosmotic flow of power law fluids in nanochannels by means of DPD simulations. Hsu et al. [70] studied the ion transport in a pH-regulated conical nanopore filled with a power-law fluid. They investigated the effects of the bulk salt concentration, the solution pH, and the power-law index on the ion current rectification (ICR) and the ion selectivity (S). In another research paper, electroosmotic flow of a LPTT fluid in a nanoslit was investigated by Mei et al. [71]. In a more recent research, Mei and Qian [72] examined the electroosmotic flow of viscoelastic fluids in a nanochannel. Linear Phan-Thien–Tanner (LPTT) constitutive model for viscoelastic fluid and Nernst−Plank equation for ionic concentration are considered. In mentioned research, enhancement of the flow rate and increase in the ionic conductance of the nanochannel as a result of shear-thinning effect are observed. In a very recent study, Barman et al. [73] investigated the electrokinetic ion transport of power-law fluids in a slit polymer-grafted nanochannel. In this research, the walls of the nanochannel are considered to be coated with the ion and fluid penetrable polymer layer containing pH-regulated functional group. To the authors best knowledge, the ionic tranport of non-Newtonian fluids in a pH-regulated rectangular nanochannel has not already been reported in literature. In the present work, a finite difference based numerical method is employed to study the ionic transport and selectivity of electrokinetically-actuated flow in a long pH-regulated rectangular nanochannel filled with power-law fluids. The effects of the channel geometry, the solution properties (e.g. background salt concentration and pH) and the non-Newtonian fluid behavior on the surface charge properties, velocity field, ionic transport, and selectivity are investigated by means of a complete parametric study. A finite element numerical study in 3D space is also performed by COMSOL Multiphysics in order to compare the results with those obtained from 2D analysis. Furthermore, an analytic solution for the limiting case of a slit geometry is presented in the end.

## 2. Problem Formulation

### 2.1. Problem definition

Consider the electroosmotic flow of a power-law fluid in a straight pH-regulated rectangular nanochannel with constant physical properties. The channel dimensions, the coordinate system located at the center of the Y-Z plane of the channel, and the other details are shown in the schematic given in Fig 1. It is assumed that length of the nanochannel is much larger than its height and width ($L \gg H, W$) and the ionic concentration polarization effect arising from the selective transport of ions through the nanochannel is neglected. The Debye length is considered to be smaller than the half height and width of the nanochannel and the overlapping of the electric double layers (EDLs) is not significant. Moreover, it is supposed that the EDLs are in equilibrium condition to allow use of Poisson-Boltzmann ionic distribution and the flow is considered to be steady, laminar, and fully developed. The effect of the Stern layer adjacent to the channel wall is also neglected. Owing to the symmetry, the analysis is limited to the first quarter of the channel. The channel wall in contact with an aqueous solution is considered to be of charge-regulated nature and bears dissociable functional groups which are capable of undergoing dissociation/association reactions with protons in aqueous solution. It is assumed that the background salt in aqueous electrolyte solution is KCl with a concentration of $C_{KCl}$ and solution pH is adjusted by KOH and HCl. Therefore, four major ionic species, namely, $K^+$, $Cl^-$, $H^+$ and $OH^-$ with $C_{i0}$, i = 1, 2, 3, and 4, being their bulk concentration (in mM) are considered. The electroneutrality of the bulk solution requires that $C_{10} = C_{KCl}$, $C_{20} = C_{KCl} + 10^{-pH+3} - 10^{-(14-p)+3}$, $C_{30} = 10^{-pH+3}$, and $C_{40} = 10^{-(14-p)+3}$ for pH ≤ 7; $C_{10} = C_{KCl} - 10^{-pH+3} + 10^{-(14-pH)+3}$, $C_{20} = C_{KCl}$, $C_{30} = 10^{-pH+3}$, and $C_{40} = 10^{-(14-p)+3}$ for pH > 7.

## 2.2. Governing equations

### 2.2.1. Electric potential field

The electrical potential in the nanochannel is described by Poisson's equation as

$$\nabla^2 \varphi = -\frac{\rho_e}{\varepsilon} \qquad (1)$$

where, $\varepsilon$ is the fluid permittivity and $\rho_e$ is the net electric charge density. The externally applied electric field is assumed low enough to allow linear combination of the externally imposed field $\Phi$ and the EDL potential $\psi$ for the overall electric field. By using the Boltzmann distribution, the net electric charge of the solution is given as

$$\rho_e = F \sum_{i=1} Z_i C_i = FZC_0 \left[ \exp\left(-\frac{eZ\psi}{K_B T}\right) - \exp\left(\frac{eZ\psi}{K_B T}\right) \right] = -2FZC_0 \sinh\left(\frac{eZ\psi}{k_B T}\right) \qquad (2)$$

where $F$ is the Faraday constant, $k_B$ shows the Boltzmann constant, $T$ indicates the absolute temperature, $Z_i$ is the valence of the $i^{th}$ ionic species, $Z = |Z_i|$ and $C_0 = C_{10} + C_{30} = C_{20} + C_{40}$.

Note that it is supposed that the solution ions obey the Boltzmann distribution since the fluid velocity does not influence the ionic distribution in the normal direction in a rectilinear flow. Substituting Equation (2) into Equation (1) and recalling that $\nabla^2 \Phi = 0$ for a constant applied electric field, one obtains

$$\frac{\partial^2 \psi}{\partial y^2} + \frac{\partial^2 \psi}{\partial z^2} = \frac{2FZC_0}{\varepsilon} \sinh\left(\frac{FZ_i \psi}{RT}\right) \qquad (3)$$

where $R$ is the universal gas constant. The above equation is subject to the following boundary conditions

$$\left.\frac{\partial \psi}{\partial y}\right|_{y=0} = \left.\frac{\partial \psi}{\partial z}\right|_{z=0} = 0 \qquad (4)$$

$$\varepsilon \left.\frac{\partial \psi}{\partial y}\right|_{y=H} = \sigma_s|_{y=H}, \quad \varepsilon \left.\frac{\partial \psi}{\partial z}\right|_{z=W} = \sigma_s|_{z=W} \qquad (5)$$

The boundary conditions (4) and (5) reflect symmetry at the centerlines and surface charge density of the channel walls, $\sigma_s$, respectively. Suppose that the channel wall bears dissociable functional groups XOH capable of experiencing the following two main dissociation/association reactions, $\text{XOH} \leftrightarrow \text{XO}^- + \text{H}^+$ and $\text{XOH} + \text{H}^+ \leftrightarrow \text{XOH}_2^+$ with equilibrium constants of $K_A = (\Gamma_{\text{XO}^-}[\text{H}^+]_s)/\Gamma_{\text{XOH}}$ and $K_B = \Gamma_{\text{XOH}_2^+}/(\Gamma_{\text{XOH}}[\text{H}^+]_s)$. Here, $\Gamma_{\text{XOH}}$, $\Gamma_{\text{XO}^-}$ and $\Gamma_{\text{XOH}_2^+}$ are surface site densities of functional group XOH, $\text{XO}^-$ and $\text{XOH}_2^+$ in sites/nm², respectively; $[\text{H}^+]_s$ is the molar concentration of $\text{H}^+$ ions at the channel liquid interface. It can be shown that surface charge density of the channel wall is expressed as [37]

$$\sigma_s = \left(\frac{-F\Gamma_t \times 10^{18}}{N_a}\right) \left(\frac{K_A - K_B\left(10^{-\text{pH}} e^{-\frac{FZ\psi_s}{RT}}\right)^2}{K_A + 10^{-\text{pH}} e^{-\frac{FZ\psi_s}{RT}} + K_B\left(10^{-\text{pH}} e^{-\frac{FZ\psi_s}{RT}}\right)^2}\right) \qquad (6)$$

where $N_a$ is the Avogadro's number, $\psi_s$ is the surface potential and $\Gamma_t = \Gamma_{\text{XOH}} + \Gamma_{\text{XO}^-} + \Gamma_{\text{XOH}_2^+}$ is the total number site density of functional groups on the surface.

By using the definition of the Debye length which is $\lambda_D = (2F^2Z^2C_0/\varepsilon RT)^{-1/2}$ and introducing the following dimensionless parameters

$$z^* = \frac{z}{H}, \quad y^* = \frac{y}{H}, \quad \psi^* = \frac{FZ\psi}{RT}, \quad K = \frac{H}{\lambda_D} \tag{7}$$

equation (3) can be non-dimensionalized as

$$\frac{\partial^2 \psi^*}{\partial y^{*2}} + \frac{\partial^2 \psi^*}{\partial z^{*2}} = K^2 \sinh \psi^* \tag{8}$$

Using Equations (4), (5), (6), and (7), the relevant boundary conditions for Equation (8) can be obtained as

$$\left.\frac{\partial \psi^*}{\partial y^*}\right|_{y^*=0} = \left.\frac{\partial \psi^*}{\partial z^*}\right|_{z^*=0} = 0 \tag{9}$$

$$\left.\frac{\partial \psi^*}{\partial y^*}\right|_{y^*=1} = \sigma_s^*|_{y^*=1}, \quad \left.\frac{\partial \psi^*}{\partial z^*}\right|_{z^*=\alpha} = \sigma_s^*|_{z^*=\alpha} \tag{10}$$

where

$$\sigma_s^* = \left(\frac{-F^2\Gamma_t ZH \times 10^{18}}{\varepsilon RTN_a}\right)\left(\frac{K_A - K_B\left(10^{-\mathrm{pH}}e^{-\psi_s^*}\right)^2}{K_A + 10^{-\mathrm{pH}}e^{-\psi_s^*} + K_B(10^{-\mathrm{pH}}e^{-\psi_s^*})^2}\right). \tag{11}$$

### 2.2.2. Velocity field

The velocity field can be obtained by solving the continuity and Cauchy momentum equations given as

$$\nabla \cdot \mathbf{u} = 0 \tag{12}$$

$$\rho \frac{D\mathbf{u}}{Dt} = -\nabla p + \nabla \cdot \boldsymbol{\tau} + \mathbf{F} \tag{13}$$

In the above equation $\mathbf{u}$ is the velocity vector, $\rho$ is the density, $p$ is the pressure, $\mathbf{F}$ is the body force vector and $\boldsymbol{\tau}$ is the stress tensor given as

$$\boldsymbol{\tau} = 2\mu(\dot{\gamma})\dot{\boldsymbol{\gamma}} \tag{14}$$

in which $\dot{\boldsymbol{\gamma}} = (\nabla \mathbf{u} + \nabla \mathbf{u}^T)/2$ is the strain rate tensor and $\mu(\dot{\gamma})$ is the effective viscosity. $\dot{\gamma}$ is the magnitude of the strain rate tensor given as [74]

$$\dot{\gamma} = \left(\frac{1}{2}\dot{\boldsymbol{\gamma}}:\dot{\boldsymbol{\gamma}}\right)^{1/2} \tag{15}$$

The effective viscosity for the power law fluids is given by [74]

$$\mu(\dot{\gamma}) = m(2\dot{\gamma})^{n-1} \tag{16}$$

where $m$ and $n$ are the flow consistency and flow behavior indexes, respectively. Since it is assumed that the flow is steady and hydrodynamically fully developed, the velocity vector becomes rectilinear and a function of the transverse coordinates as $\mathbf{u} = [u(y,z),0,0]$. So, the following equations for $\dot{\gamma}$ and $\mu$ are obtained

$$\dot{\gamma} = \frac{1}{2}\left[\left(\frac{\partial u}{\partial y}\right)^2 + \left(\frac{\partial u}{\partial z}\right)^2\right]^{1/2} \tag{17}$$

$$\mu(\dot{\gamma}) = m\left[\left(\frac{\partial u}{\partial y}\right)^2 + \left(\frac{\partial u}{\partial z}\right)^2\right]^{\frac{n-1}{2}} \tag{18}$$

Using Equations (14) and (18), the following relations for shear stress components in the axial direction can be obtained

$$\tau_{yx} = m\left[\left(\frac{\partial u}{\partial y}\right)^2 + \left(\frac{\partial u}{\partial z}\right)^2\right]^{\frac{n-1}{2}} \frac{\partial u}{\partial y} \tag{19}$$

$$\tau_{zx} = m\left[\left(\frac{\partial u}{\partial y}\right)^2 + \left(\frac{\partial u}{\partial z}\right)^2\right]^{\frac{n-1}{2}} \frac{\partial u}{\partial z} \tag{20}$$

Since, it is assumed that the flow is electroosmotically driven, there is no pressure gradient and gravitational body force. The axial component of the electric body force due to applied electric field equals $\rho_e E_x$ with $E_x = -d\Phi/dx$ is the electric field in the axial direction. Therefore, utilizing Equation (2), the axial component of the electric body force becomes

$$F_x = -2FZC_0E_x\sinh\left(\frac{FZ\psi}{RT}\right) \tag{21}$$

Substituting Equations (19), (20), and (21) in Equation (13), the Cauchy momentum equation for a steady fully developed flow in the axial direction becomes

$$\frac{\partial}{\partial y}\left\{m\left[\left(\frac{\partial u}{\partial y}\right)^2 + \left(\frac{\partial u}{\partial z}\right)^2\right]^{\frac{n-1}{2}}\frac{\partial u}{\partial y}\right\} + \frac{\partial}{\partial z}\left\{m\left[\left(\frac{\partial u}{\partial y}\right)^2 + \left(\frac{\partial u}{\partial z}\right)^2\right]^{\frac{n-1}{2}}\frac{\partial u}{\partial z}\right\}$$
$$- 2FZC_0E_x\sinh\left(\frac{FZ\psi}{RT}\right) = 0 \tag{22}$$

By expanding the first two terms of equation (22) and by defining $u^* = u/\mathbb{U}$, where $\mathbb{U} = n\lambda_D^{\frac{n-1}{n}}\left(-\frac{\varepsilon RTE_x}{FZm}\right)^{\frac{1}{n}}$, the axial component of the momentum equation takes the following form

$$A_1(y^*,z^*,u^*)\frac{\partial^2 u^*}{\partial y^{*2}} + A_2(y^*,z^*,u^*)\frac{\partial^2 u^*}{\partial z^{*2}} + A_3(y^*,z^*,u^*)\frac{\partial^2 u^*}{\partial y^* \partial z^*} = -\frac{K^{n+1}}{n^n}\sinh\psi^* \tag{23}$$

where $A_1(y^*,z^*,u^*)$, $A_2(y^*,z^*,u^*)$ and $A_3(y^*,z^*,u^*)$ are non-dimensional functions given as

$$A_1(y^*,z^*,u^*) = \left[\left(\frac{\partial u^*}{\partial y^*}\right)^2 + \left(\frac{\partial u^*}{\partial z^*}\right)^2\right]^{\frac{n-3}{2}}\left[n\left(\frac{\partial u^*}{\partial y^*}\right)^2 + \left(\frac{\partial u^*}{\partial z^*}\right)^2\right]$$

$$A_2(y^*,z^*,u^*) = \left[\left(\frac{\partial u^*}{\partial y^*}\right)^2 + \left(\frac{\partial u^*}{\partial z^*}\right)^2\right]^{\frac{n-3}{2}}\left[\left(\frac{\partial u^*}{\partial y^*}\right)^2 + n\left(\frac{\partial u^*}{\partial z^*}\right)^2\right] \tag{24}$$

$$A_3(y^*,z^*,u^*) = 2(n-1)\left[\left(\frac{\partial u^*}{\partial y^*}\right)^2 + \left(\frac{\partial u^*}{\partial z^*}\right)^2\right]^{\frac{n-3}{2}}\frac{\partial u^*}{\partial y^*}\frac{\partial u^*}{\partial z^*}$$

The dimensionless momentum equation is subject to the no slip conditions at the wall and the symmetry conditions at the channel midplanes given as

$$\left.\frac{\partial u^*}{\partial y^*}\right|_{y^*=0} = \left.\frac{\partial u^*}{\partial z^*}\right|_{z^*=0} = 0 \tag{25}$$

$$u^*|_{y^*=1} = u^*|_{z^*=\alpha} = 0. \tag{26}$$

## 2.3. Parameters of interest

One of the parameters of physical interest is the average of the surface potential over the entire surfaces of the nanochannel which can be expressed as

$$\psi_{s,av} = \frac{RT}{FZ}\left[\int_0^\alpha \psi_s^*|_{y^*=1}\, dz^* + \int_0^1 \psi_s^*|_{z^*=\alpha}\, dy^*\right] \quad (27)$$

The surface charge density plays a key role in determining the ionic conductance and flow parameters of the nanochannel, so, by invoking Equation (5), the average of the surface charge density at the channel wall can be evaluated as

$$\sigma_{s,av} = \frac{1}{1+\alpha}\left(\frac{\varepsilon RT}{FZH}\right)\left[\int_0^\alpha \sigma_s^*|_{y^*=1}\, dz^* + \int_0^1 \sigma_s^*|_{z^*=\alpha}\, dy^*\right] \quad (28)$$

where $\sigma_s^*$ is given by Equation (11). In addition, by using Equation (2), the average electric charge in the solution can be obtained as

$$\rho_{e,av} = \frac{1}{\alpha}\int_0^\alpha \int_0^1 \rho_e\, dy^* dz^* = \frac{1}{\alpha}\int_0^\alpha \int_0^1 -2FZC_0 \sinh(\psi^*)\, dy^* dz^* \quad (29)$$

Another parameter of interest is the mean velocity which is given as

$$u_m = \frac{1}{\alpha}\mathbb{U}\int_0^\alpha \int_0^1 u^*\, dy^* dz^* \quad (30)$$

The average of viscosity at the channel surface can also be expressed as

$$\mu_{s,av} = \frac{1}{1+\alpha}\left[\int_0^\alpha \mu|_{y^*=1}\, dz^* + \int_0^1 \mu|_{z^*=\alpha}\, dy^*\right]$$
$$= \frac{m^{1/n}n^{n-1}}{(1+\alpha)k^{n-1}}\left[-\frac{\varepsilon RT E_x}{\lambda_D FZ}\right]^{\frac{n-1}{n}}\left\{\int_0^\alpha \left(\left(\frac{\partial u^*}{\partial y^*}\right)^2\right)^{\frac{n-1}{2}}\bigg|_{y^*=1} dz^*\right.$$
$$\left. + \int_0^1 \left(\left(\frac{\partial u^*}{\partial z^*}\right)^2\right)^{\frac{n-1}{2}}\bigg|_{z^*=\alpha} dy^*\right\} \quad (31)$$

The ionic flux in the nanochannel consists of bulk convection, ionic diffusion and migration of ions. Since an equilibrium distribution for the ionic species is considered and it is assumed that the nanochannel is long enough that the concentration polarization effect could be neglected, the variation of the ionic concentration in the axial direction is insignificant. Therefore, the ionic flux will be the sum of the migratory and convective fluxes which can be expressed as

$$N_{x,i} = \frac{D_i Z_i F}{RT} C_i E_x + C_i u \tag{32}$$

where, $N_{x,i}$, $D_i$ and $Z_i$ are the axial flux, diffusivity and the valence of the i$^{th}$ ionic species, respectively. The mean current density can be obtained as

$$J = \frac{1}{\alpha} \int_0^\alpha \int_0^1 F \left( \sum_i Z_i N_{x,i} \right) dy^* dz^* \tag{33}$$

from which the ionic current can be evaluated as $I = 4\alpha H^2 J$. By substituting Equation (32) in Equation (33) and by assuming Boltzmann distribution, the following expressions for the convective and migratory parts of the current density ($J_c$ and $J_m$) can be obtained

$$J_c = \frac{-2FZC_0}{\alpha} \mathbb{U} \int_0^\alpha \int_0^1 u^* \sinh(\psi^*) dy^* dz^* \tag{34}$$

$$J_m = \frac{F^2 Z^2 E_x}{\alpha RT} \int_0^\alpha \int_0^1 \{(C_{10}D_1 + C_{30}D_3)e^{-\psi^*} + (C_{20}D_2 + C_{40}D_4)e^{\psi^*}\} dy^* dz^* \tag{35}$$

Once the ionic current is obtained, the ionic conductance contributions can be evaluated by $G = I/V$ as the following

$$G_c = \frac{-8FZC_0 H^2}{V} \mathbb{U} \int_0^\alpha \int_0^1 u^* \sinh(\psi^*) dy^* dz^* \tag{36}$$

$$G_m = \frac{4F^2 Z^2 E_x H^2}{RTV} \int_0^\alpha \int_0^1 \{(C_{10}D_1 + C_{30}D_3)e^{-\psi^*} + (C_{20}D_2 + C_{40}D_4)e^{\psi^*}\} dy^* dz^*, \tag{37}$$

where $V$ is the voltage difference between the channel ends. The ionic selectivity of the nanochannel can also be expressed as

$$S = \frac{|I_{ca}| - |I_{an}|}{|I_{ca}| + |I_{an}|} \tag{38}$$

in which $I_{ca}$ and $I_{an}$ can be evaluated from the following expressions

$$I_{ca} = I_1 + I_3 = 4FZC_0H^2\mathbb{U}\int_0^\alpha\int_0^1 e^{-\psi^*}u^*dy^*dz^*$$
$$+ \frac{4F^2Z^2E_xH^2}{RT}\int_0^\alpha\int_0^1 (C_{10}D_1 + C_{30}D_3)e^{-\psi^*}dy^*dz^*, \quad (39)$$

$$I_{an} = I_2 + I_4 = -4FZC_0H^2\mathbb{U}\int_0^\alpha\int_0^1 e^{\psi^*}u^*dy^*dz^*$$
$$+ \frac{4F^2Z^2E_xH^2}{RT}\int_0^\alpha\int_0^1 (C_{20}D_2 + C_{40}D_4)e^{\psi^*}dy^*dz^* \quad (40)$$

In the above equations, the first and the second parts are related to convective and migratory currents respectively.

### 2.4. 3D analysis

In order to relax the assumption made in derivation of the governing equations in 2D forms, in this section the numerical analysis of the problem in 3D space will be performed using COMSOL Multiphysics software. The Cauchy momentum equation, assuming low values for Reynolds number can be expressed as

$$-\nabla p + \nabla \cdot \left\{m(2\dot{\boldsymbol{\gamma}}:\dot{\boldsymbol{\gamma}})^{\frac{n-1}{2}}(\nabla \mathbf{u} + \nabla \mathbf{u}^T)\right\} + \mathbf{F} = 0 \quad (41)$$

Therefore, the non-dimensional governing equations in the following forms are obtained

$$\nabla^{*2}\Phi^* = 0 \quad (42)$$

$$\nabla^{*2}\psi^* = K^2 \sinh\psi^* \quad (43)$$

$$\nabla^* \cdot \mathbf{u}^* = \mathbf{0} \quad (44)$$

$$-\nabla^*p^* + \nabla^* \cdot \left\{(2\dot{\boldsymbol{\gamma}}^*:\dot{\boldsymbol{\gamma}}^*)^{\frac{n-1}{2}}(\nabla^*\mathbf{u}^* + \nabla^*\mathbf{u}^{*T})\right\} + 2\sinh\psi^*\nabla^*(\Phi^* + \psi^*) = 0 \quad (45)$$

where, new parameters are defined as

$$p^* = \frac{p}{C_0RT}, \mathbf{u}^* = \frac{\mathbf{u}}{H\left(\frac{C_0RT}{m}\right)^{1/n}}, \Phi^* = \frac{FZ\Phi}{RT}, \psi^* = \frac{FZ\psi}{RT}, \nabla^* = H\nabla \quad (46)$$

The relevant boundary conditions for solving Equations (42) to (45) are given in Table 1. The computations were performed by utilizing a total number of 1090618 tetrahedral meshes, assuming L = 30H in the physical domain given by Figure 1, and using finer meshes at inlet, outlet and near the walls of the nanochannel in order to provide sufficiently accurate and grid-independent results.

### 3. Results and Discussions

In order to study the electroosmotic flow of power-law fluids in nanochannels, Equations (8) and (23) subject to relevant boundary conditions are numerically solved through a finite difference procedure for a non-uniform grid. The details of method can be found in our previous study [58] and it is not presented here in order to save space. Unless otherwise stated, the values of the physicochemical parameters are set according to Table 2. To ensure the grid independency of the results, a grid dependency analysis of the convective ionic conductance and ionic conductance values is performed. As observed in Table 3, there is not any considerable change in the results when the grid system 200×200 is replaced by 250 × 250. In fact, the maximum difference between the results obtained utilizing these grid systems is below 0.01%. Therefore, it can be concluded that using 200 × 200 grid points results in adequately grid-independent results. Furthermore, results from numerical simulation are also compared with those obtained from analytical solution under Debye-Hückel approximation for a limiting case of a slit nanochannel. A detailed discussion on derivation of the analytical solution along with the comparison of the results are provided in the appendix.

The analysis starts with Figure 2 which illustrates the velocity distribution for two different values of the flow behavior index and solution pH. It should be pointed out that since the reference velocity used for non-dimensionalization of the velocity is a function of the background salt concentration and flow behavior index, the results presented here are converted to the dimensional ones in order to have a reasonable comparison. As can be seen, by lowering the pH value from 3 to 2, the sign of the velocity changes from positive to negative. This is due to the fact that the sign of the surface charge is dependent on the deviation of the pH value from the isoelectric point (a point in which the positive and negative surface charges are equal, which corresponds to 2.55 here). While the sign of the surface charge is negative for pH=3, it is positive for pH=2, leading to the electric body forces in opposite directions and consequently different sign of the velocities. In addition, it is observed that the shear-thinning fluids reach higher velocities in comparison with the shear-thickenings. The reason is that a shear-thinning fluid presents smaller viscosities at the wall [53], resulting in a lower resistance to flow and a higher velocity. It should be mentioned that for more visibility only $25 \times 25$ grid points have been utilized to sketch these figures.

The variation of the average surface potential versus the background salt concentration is depicted in Figure 3 at two different values of the channel aspect ratio and various values of the solution pH. It is shown that the magnitude of the average surface potential is a decreasing function of the background salt concentration. By elevating the background salt concentration, EDL is more and more confined to the surface. Therefore, gathering more counterions near the wall will result in the reduction of the magnitude of the potential near the wall region as well as the wall surface. It is also apparent that the magnitude of $\psi_{s,av}$ is higher for smaller channel aspect and the difference becomes less as $C_{KCl}$ increases. The former is due to the higher effect of vertical walls on potential distribution near the horizontal walls at smaller aspect ratio and the latter is due to the fact that by increasing $C_{KCl}$, EDL gets thinner and more limited to the area near the channel walls, leading to the decrement of channel aspect ratio effect.

Figure 4 displays the variation of the average surface charge density versus the background salt concentration at various values of the solution pH and two different values of the channel aspect ratio. It is observed that by increasing $C_{KCl}$, the magnitude of $\sigma_{s,av}$ is also increased. When pH is above the IEP, the surface charge is negative and both $K^+$ and $H^+$ ions are attracted by the charged surface. By raising $C_{KCl}$, concentration of $K^+$ will increase in the entire solution as well as near the wall region. Furthermore, EDL is more and more confined to the wall region and elevation in $K^+$ ions near the wall will lead to replacement of $H^+$ ions with $K^+$ ions, favoring the dissociation reaction to the forward direction based on the Le Chatelier's principle, and consequently increasing the magnitude of $\sigma_{s,av}$. When pH is lower than IEP, the surface charge is positive and both $K^+$ and $H^+$ ions are repelled from the surface. By raising $C_{KCl}$, as was explained before, surface potential decreases which results in less repulsion of $H^+$ ions from the surface. Therefore, elevation of $H^+$ concentration near the wall region will cause the magnitude of $\sigma_{s,av}$ to increase by favoring the association reaction in the forward direction. It is also apparent that the magnitude of $\sigma_{s,av}$ is smaller for lower value of the channel aspect ratio and the difference becomes less as $C_{KCl}$ increases because of thinner EDLs. For smaller aspect ratio, as was observed before, the magnitude of $\psi_{s,av}$ is greater which leads to the more attraction/repulsion of $H^+$ to the wall region for pH values greater/smaller than IEP and consequently reduction in magnitude of $\sigma_{s,av}$.

In Figure 5, the velocity profile along the channel height at the channel center, that is at $z = 0$, is compared with that obtained by Comsol Multiphysics. The comparison for two different values of the flow behavior index and background salt concentration reveals good agreements between the results. As expected, the velocity for $n = 0.9$ is higher than that for $n = 1.1$. Furthermore, as can be seen, the velocity profile becomes more plug-like by increasing the background salt concentration. This is because by increasing the background salt concentration, EDL is more concentrated near the wall region and creates a more plug-like profile. As can be seen, by enhancing $C_{KCl}$, although the electric body force increases due to the enhancement of surface charge density (as depicted in Figure 4), the velocity decreases. This can be attributed to the fact that the viscosity force should be in balance with the electroosmotic volumetric force; however, the increase in the body force with increasing $C_{KCl}$ is so small that is not even sufficient to overcome the extra wall shear rate due to the shrinkage of the EDL. Hence, the imbalance in the electroosmotic and viscous forces is compensated for by a decrease in the velocity.

The dependency of $\rho_{e,av}$ on the background salt concentration at various values of the solution pH and two different values of the channel aspect ratio is illustrated in Figure 6. Regardless of the solution pH and channel aspect ratio, the magnitude of $\rho_{e,av}$ is seen to have an increasing dependency of the background salt concentration because of higher ionic strength in the solution. It is also clear that the magnitude of $\rho_{e,av}$ increases by the deviation of the solution pH from IEP because of higher association/dissociation reactions at the wall which leads to higher surface charges and higher attraction of counter-ions inside the channel. Moreover, it is revealed that the magnitude of $\rho_{e,av}$ is greater for smaller channel aspect ratio. This is because for channels with smaller aspect ratio, the area occupied by EDL is higher, leading to the increment of $\rho_{e,av}$. It should be noted that the total electric charge ($\rho_t$) is an increasing function of the channel aspect ratio because the channel cross section is greater for larger $\alpha$. Hence, total electric charge should not be confused by the average electric charge in the solution.

The variation of average fluid viscosity at the wall with $C_{KCl}$ for four different pH values is shown in Figure 7. As is visible, by increasing the background salt concentration or deviation of the solution pH from IEP, the viscosity decreases for a shear thinning fluid while it increases for a shear thickening type. This could be justified as by magnifying the background salt concentration or pH deviation from the isoelectric point, the electric body force enhances, leading to the increment of the shear rate at the wall. This will result in a higher viscosity for a shear thickening fluid and a lower one for a shear thinning type.

The influence of the background salt concentration on the mean velocity at various values of the flow behavior index and channel aspect ratio is investigated in Figure 8. As can be seen, the mean velocity is a decreasing function of the background salt concentration. This is because by increasing the background salt concentration, the EDL and velocity variation will be confined to a thin area near the wall, leading to higher shear rates at the wall and consequently smaller velocities. Furthermore, the mean velocity is found to be higher for greater channel aspect ratios. The reason is that by increasing the channel aspect ratio, the vertical wall influence and its resistance to flow as a result of smaller values of the shear rate at the wall will decrease, resulting in a higher mean velocity.

The dependency of the mean velocity on the flow behavior index at various values of the solution pH is illustrated in Figure 9. It is observed that $|u_m|$ decreases by increasing the flow behavior index because of higher viscosities at the wall. Furthermore, $|u_m|$ increases by the deviation of the solution pH from IEP because of higher net electric charge and body force in the solution. The point drawing attention in this figure is that while the deviation of the solution pH from IEP is equal for pH=1 and pH=4.1, the magnitude of $u_m$ is greater for pH=4.1. This is due to the fact that for pH=1, $H^+$ concentration is significant in the solution. Thus, the resulted thinner EDL will lead to higher shear rates at the wall and consequently smaller values of $|u_m|$.

Moreover, for higher values of n, because the friction forces get dominant, the fluid velocity tends to zero for all pH values.

The variation of the convective ionic conductance versus the background salt concentration for various values of the flow behavior index and two different values of the channel aspect ratio is disclosed in Figure 10. It is observed that by increasing the background salt concentration or the flow behavior index, the convective ionic conductance decreases. This is because $G_c$ is directly proportional to the fluid velocity and the velocity decreases by an increase in $C_{KCl}$ or n. It is also clear that greater aspect ratio will result in a higher convective ionic conductance which is due to the higher velocity magnitude for greater channel aspect ratios. In this figure the ionic conductance multiplied by 30 is also compared with the experimental data reported by Karnik et al. [75] for an array of 30 silica nanochannles, revealing an acceptable agreement between the results.

Figure 11 depicts the variation of the ionic conductance versus the solution pH for various values of the flow behavior index and channel aspect ratio. It is evident that the effect of non-Newtonian fluid behavior becomes remarkable when the values of pH exceed 5 and the ionic conductance is higher for the shear-thinning fluids compared to the shear-thickenings. This is due to the fact that at higher values of the solution pH, because of higher surface charges, the effect of the convective ionic conductance becomes more significant and as was observed before, $G_c$ is a decreasing function of the flow behavior index. As was expected, ionic conductance increases by the deviation of the solution pH from IEP because of higher total electric charge in the solution. It is also observed that the ionic conductance is higher for a channel with greater aspect ratio because of higher total electric charge therein.

The dependency of the ionic selectivity on the background salt concentration is displayed in Figure 12. For the considered value of pH = 4.5 and low values of $C_{KCl}$, the nanochannel is cation selective ($S > 0$) because pH is greater than IEP and $K^+$ and $H^+$ ions mostly occupy the nanochannel environment and contribute to the nanochannel conductance. As observed, by increasing $C_{KCl}$, the ion selectivity decreases because EDL becomes more and more confined to the wall region, causing less resistance for the co-ions to enter inside the nanochannel. Furthermore, it can be seen that the ion selectivity is greater for smaller channel aspect ratio because greater portion of the channel cross section is occupied by EDL for smaller $\alpha$. It is also apparent that the ion selectivity is greater for $n = 0.9$ in comparison with $n = 1.1$. This is rooted in the fact that as can be seen in part b of this figure, $I_{ca}$ is greater than $I_{an}$ for $n = 0.9$, while the opposite is true for $n = 1.1$. This leads to the increment of the numerator in Equation (38) and increasing the parameter $S$ for $n = 0.9$. It should also be noted that the reason of increasing $I_{ca}$ with decreasing n in part (b) is due to the enhancement of fluid velocity in favor of counterions movement toward cathode. Similarly, this trend results in the reduction of anionic current because of suppression of co-ions transport toward the anode electrode and as a result decrement of $I_{an}$.

The variation of the ionic selectivity versus pH is illustrated in Figure 13. It can be seen that by increasing pH from low values, selectivity declines because decrement in $H^+$ ions leads to a sharp decrease in migratory current from $H^+$ ions in the solution. By further elevation in pH values, selectivity begins to increase until it reaches its maximum value of unity for both cases of $n = 0.9$ and 1. This happens because for pH values higher than IEP, raising pH leads to a rise in fluid velocity, which causes $I_{an}$ to decrease as mentioned before. Thereby, when $I_{an}$ gets close to zero at a threshold value of pH, selectivity reaches its apex based on Equation (38). By raising pH over this threshold point, $I_{an}$ becomes negative and therefore selectivity starts to decline until it reaches its local minima. One point which draws attention in this figure is that the threshold pH value corresponds to the selectivity maxima for case of $n = 0.9$ is lower than $n = 1$. This can be attributed to the vanishment of $I_{an}$ at lower pH values for $n = 0.9$ as depicted in part (b). It should also be noted that by decreasing the flow behavior index, the convective part of the anionic transport can become higher than the migration contribution, especially for high values of solution pH. This trend is arisen from higher velocity field inside the nanochannel.

## 4. Conclusions

In this study, the ionic transport and selectivity of a fully developed electrokinetically-driven flow of power-law fluids in a long pH-regulated rectangular nanochannel were analyzed. Having been made dimensionless, the electrical potential and momentum equations were numerically solved through a finite difference procedure for a non-uniform grid. The results were compared with those obtained from a full 3D numerical simulation using COMSOL Multiphysics software as well as those evaluated from the derived analytical expressions. A thoroughgoing parametric study revealed that the channel aspect ratio and the non-Newtonian characteristic of the fluid can influence all the physicochemical parameters. It is found that as the channel approaches a square shape, the magnitude of the average surface potential increases while the opposite trend was observed for the magnitude of the average surface charge density. The dependency of both mentioned parameters on the channel aspect ratio diminishes sharply at high values of the background salt concentration. The magnitude of the average electric charge in the solution is found to be higher for smaller channel aspect ratio for the entire range of the background salt concentration. The mean velocity and the convective ionic conductance are observed to be decreasing functions of the flow behavior index and mentioned parameters are higher for greater values of the channel aspect ratio. It is also revealed that the ionic conductance is strongly dependent on flow behavior index at high pH values and it is higher for shear-thinning fluids in comparison with shear-thickenings. Furthermore, it is shown that the ion selectivity of the nanochannel can strongly be affected by the flow behavior index, background salt concentration, and solution pH and it is found that the variation of the flow behavior index can be utilized to let the selectivity go through its maximum as a function of pH.

**Notes**

The authors declare no conflicts of interest.

**Data Availability**

The data of this study are available from the corresponding author upon reasonable request.

**Appendix A**

In some of the situations, the channel aspect ratio is large enough that the electrical potential and velocity fields are not significantly changed over the majority of the channel width. The simpler and less expensive method to treat the problem in these situations is to approximate the rectangular geometry by two parallel plates with a distance H apart. Therefore, in this section the analytical expressions of the potential and velocity profiles for electroosmotic flow of power-law fluids in a slit nanochannel are derived for the case of low surface potential. To conduct the analysis, we may begin to find the electrical potential distribution in the nanochannel using the Poisson-Boltzmann equation along with the Debye-Hückel linearization as

$$\frac{d^2\psi}{dy^2} = \lambda_D^{-2}\psi \tag{A1}$$

in which, $\lambda_D$ is the Debye length parameter defined as in section 2.2.1. Considering the association/dissociation reactions on the surface, one can obtain the boundary conditions associated to the above equation as below

$$\varepsilon \frac{d\psi}{dy}\bigg|_{y=H} + \Xi\psi|_{y=H} = \Omega \tag{A2}$$

$$\frac{d\psi}{dy}\bigg|_{y=0} = 0 \tag{A3}$$

where, $\Xi$ and $\Omega$ are given as [46]

$$\Omega = \sigma_0 \frac{\delta\sinh\Psi_N}{1+\delta\cosh\Psi_N}, \quad \Xi = \frac{\sigma_0}{\Psi_0} \frac{\delta\cosh\Psi_N + \delta^2}{(1+\delta\cosh\Psi_N)^2} \tag{A4}$$

in which, $\sigma_0 = F\Gamma_s$, $\Psi_0 = RT/F$, $\Psi_N = (pH_0 - pH)\ln 10$ with $pH_0 = (pK_A - pK_B)/2$, $\delta = 2 \times 10^{-\Delta pK/2} = 2\sqrt{K_A K_B}$, and $\Delta pK = pK_A + pK_B$ where $\Gamma_s = 10^{18}\Gamma_t/N_A$ is the molar site density of the functional groups [46].

By introducing the following new dimensionless parameters

$$\xi = \frac{\Xi H}{\varepsilon}, \quad \Lambda = \frac{\Omega F}{RT\Xi}, \quad \psi^* = \frac{\psi F}{RT}, \quad \varphi = \psi^* - \Lambda \tag{A5}$$

Equations (A1) – (A3) can be non-dimensionalized as

$$\frac{d^2\varphi}{dy^{*2}} = \mathbb{K}^2(\varphi + \Lambda) \tag{A6}$$

$$\frac{d\varphi}{dy^*}\bigg|_{y^*=1} + \xi\varphi_{y^*=1} = 0 \tag{A7}$$

$$\frac{d\varphi}{dy^*}\bigg|_{y^*=0} = 0 \tag{A8}$$

The analytical solution satisfying Equations (A6)-(A8) can be obtained as

$$\varphi = \frac{\xi\Lambda\cosh(\mathbb{K}y^*)}{\mathbb{K}\sinh(\mathbb{K}) + \xi\cosh(\mathbb{K})} - \Lambda \tag{A9}$$

Therefore, the electrical potential within the nanochannel is given as

$$\psi^*(y^*) = \varphi(y^*) + \Lambda = \frac{\xi\Lambda\cosh(\mathbb{K}y^*)}{\mathbb{K}\sinh(\mathbb{K}) + \xi\cosh(\mathbb{K})} \tag{A10}$$

To find the velocity profile, the Cauchy momentum equation must be solved. For a slit nanochannel filled with a power-law fluid, this equation can be simplified as

$$\frac{d}{dy}\left[m\left(-\frac{du}{dy}\right)^{n-1}\frac{du}{dy}\right] + F_x = 0 \tag{A11}$$

where $F_x$ is the electric body force equals $\rho_e E_x$ and can be written as

$$F_x = -\varepsilon\nabla^2\psi E_x = -\frac{\Omega}{H}\frac{\mathbb{K}^2\cosh(\mathbb{K}y^*)}{\mathbb{K}\sinh(\mathbb{K}) + \xi\cosh(\mathbb{K})} \tag{A12}$$

Substituting Eq. (A12) in Eq. (A11), and defining the new dimensionless velocity parameter $u^* = u/\mathbb{U}_\parallel$, where $\mathbb{U}_\parallel = H(-\Omega E_x/m)^{1/n}$, the momentum equation becomes

$$\frac{d}{dy^*}\left[\left(-\frac{du^*}{dy^*}\right)^n\right] - \frac{\mathbb{K}^2\cosh(\mathbb{K}y^*)}{\mathbb{K}\sinh(\mathbb{K}) + \xi\cosh(\mathbb{K})} = 0 \tag{A13}$$

By integrating Eq. (A13) twice, the velocity profile within the nanochannel is obtained as

$$u(y^*) = \mathbb{U}_\parallel n\left(\frac{\mathbb{K}^{1-n}}{\mathbb{K}\sinh(\mathbb{K}) + \xi\cosh(\mathbb{K})}\right)^{1/n}[G(n,\mathrm{K}) - G(n,\mathrm{K}y^*)] \tag{A14}$$

where $G(n,\omega)$ is a two-variable function defined as

$$G(n,\omega) = \frac{(-1)^{\frac{n-1}{2n}}}{n}\cosh(\omega)\,_2F_1\left[\frac{1}{2},\frac{n-1}{2n};\frac{3}{2};\cosh^2(\omega)\right], \tag{A15}$$

which $_2F_1[a, b; c, d]$ denotes the hypergeometric function.

Velocity profile in a slit nanochannel for two different values of flow behavior index and pH is illustrated in Figure A1. In this figure, the velocity distribution obtained from the numerical method described in section 2.2 for a rectangular nanochannel is also displayed. The aspect ratio of the rectangular nanochannel is fixed to a high value of $\alpha = 10$. As it is clear, for pH=3, there is a complete agreement between the results; However, when pH is magnified to 4, the difference between the results becomes a bit greater. This is due to the fact that by elevating pH value, the magnitude of the surface potential increases and the Debye-Hückel approximation for low surface potentials is no longer accurate. Therefore, this leads to a bit higher discrepancy between the analytical and numerical results obtained for pH=4.

**Table 1**. Boundary conditions associated with the governing equations solved by COMSOL Multiphysics

| Boundary | Governing Equation | | |
| --- | --- | --- | --- |
| | Equation 42 | Equation 43 | Equation 44 and 45 |
| Inlet of the nanochannel | $\Phi^* = V_{an}^*$ | $\psi^* = 0$ | $p^* = 0$ |
| Outlet of the nanochannel | $\Phi^* = 0$ | $\psi^* = 0$ | $p^* = 0$ |
| Nanochannel wall | $n . \nabla^* \Phi^* = 0$ | $n . \nabla^* \psi^* = -\sigma_s^*$ | No Slip |
| Half surface of the nanochannel $(y^*, z^* = 0)$ | $n . \nabla^* \Phi^* = 0$ | $n . \nabla^* \psi^* = 0$ | Symmetry |

**Table 2.** Values of physicochemical parameters along with the dimensional geometry used in the modeling

| Parameter | Value | parameter | Value |
| --- | --- | --- | --- |
| $m$ (pa s$^n$) | $0.9 \times 10^{-3}$ | $D_i \times 10^9 (\text{m}^2 \text{ s}^{-1})$ $i = K^+, Cl^-, H^+, OH^-$ | 1.96, 2.03, 9.31, 5.3 |
| $F$ (C mol$^{-1}$) | 96485.33 | $\mathcal{Z}_i, \ i = K^+, Cl^-, H^+, OH^-$ | $1, -1, 1, -1$ |
| $R$ (J mol$^{-1}$K$^{-1}$) | 8.314 | $\varepsilon (= \varepsilon_r \varepsilon_0)$ F m$^{-1}$ | $80 \times 8.854 \times 10^{-12}$ |
| $T$ (K) | 300 | $H$ (nm) | 40 |
| $K_A$ | $10^{-7}$ | $L = 30H$ (µm) | 1.2 |
| $K_B$ | $10^{-1.9}$ | $V_{an}$(V) | 0.03 |
| $\Gamma_t$ (sites nm$^{-2}$) | 8 | $E_x = V_{an}/L$ (V m$^{-1}$) | $25 \times 10^3$ |
| $N_a$ (mol$^{-1}$) | $6.02214 \times 10^{23}$ | | |

**Table 3.** Grid dependency of the convective ionic conductance and ionic conductance values, considering $\alpha = 2$, and pH = 5.

| Number of grid points | $C_{KCl} = 0.01$M | | | | $C_{KCl} = 0.1$M | | | |
|---|---|---|---|---|---|---|---|---|
| | $n = 0.9$ | | $n = 1.1$ | | $n = 0.9$ | | $n = 1.1$ | |
| | $G_c$ (pS) | $G$ (nS) | $G_c$ (pS) | $G$ (nS) | $G_c$ (pS) | $G$ (nS) | $G_c$ (pS) | $G$ (nS) |
| 50 × 50 | 63.1252 | 1.69034 | 6.30882 | 1.63353 | 53.4830 | 15.9668 | 4.88996 | 15.9182 |
| 100 × 100 | 63.1758 | 1.69043 | 6.31306 | 1.63356 | 53.5613 | 15.9669 | 4.89604 | 15.9182 |
| 150 × 150 | 63.1845 | 1.69044 | 6.31378 | 1.63357 | 53.5739 | 15.9669 | 4.89702 | 15.9182 |
| 200 × 200 | 63.1874 | 1.69044 | 6.31402 | 1.63357 | 53.5781 | 15.9669 | 4.89734 | 15.9182 |
| 250 × 250 | 63.1887 | 1.69045 | 6.31413 | 1.63357 | 53.5800 | 15.9669 | 4.89748 | 15.9182 |

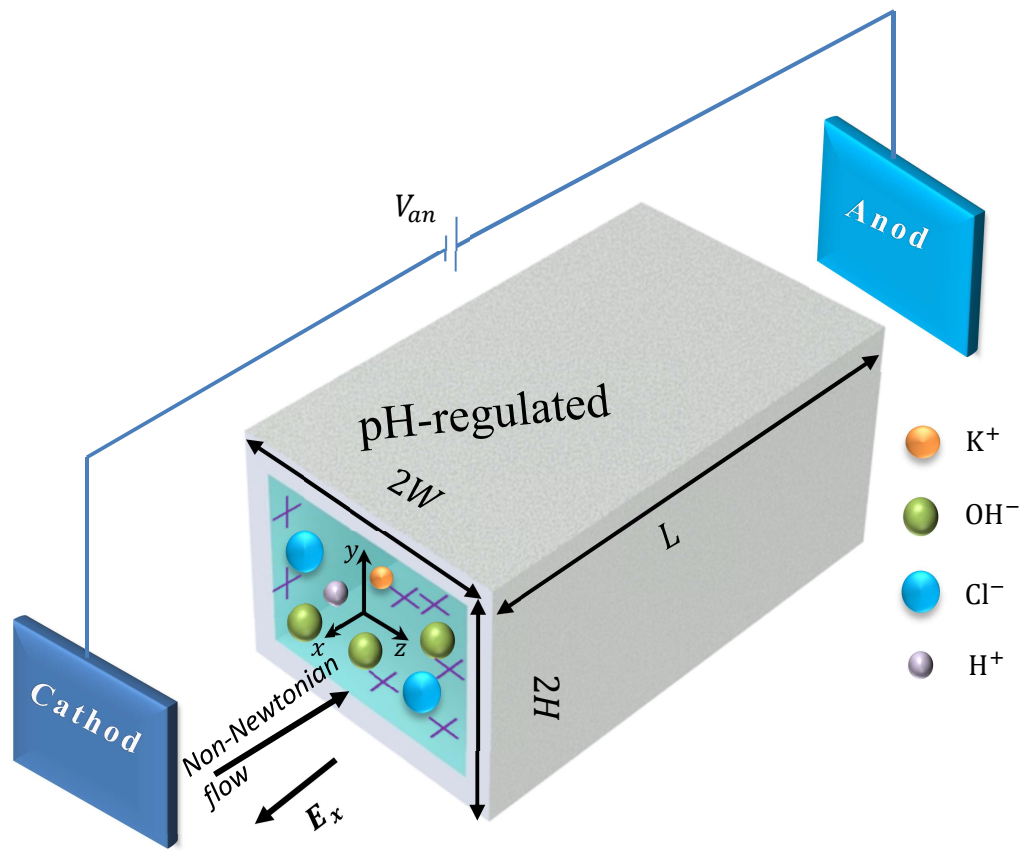

**Figure 1**. Schematic of the physical problem along with the coordinate system.

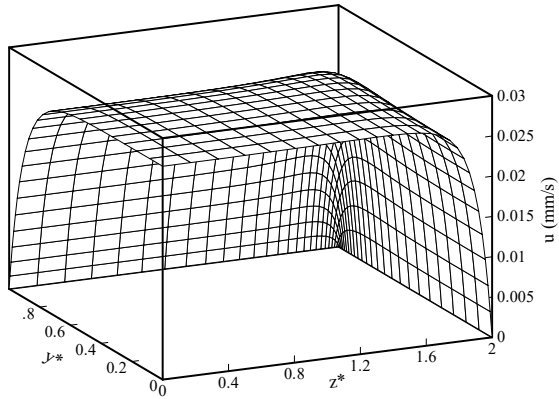 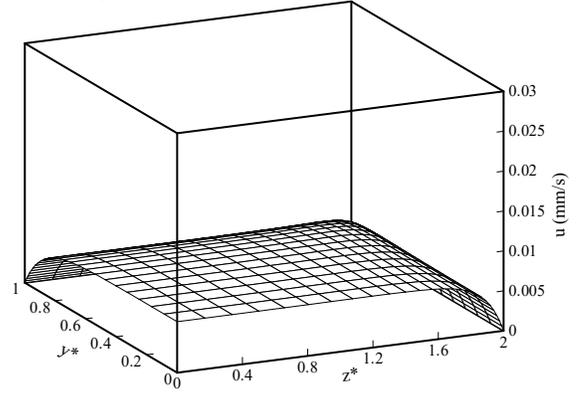
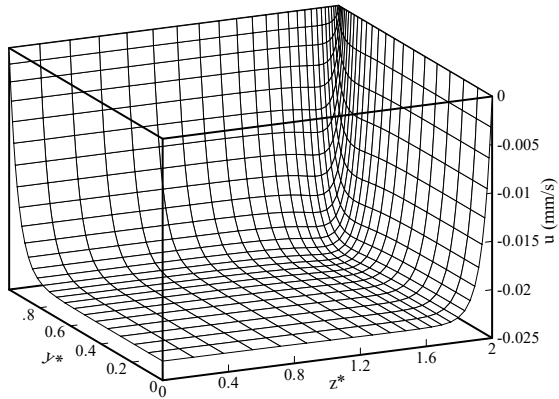 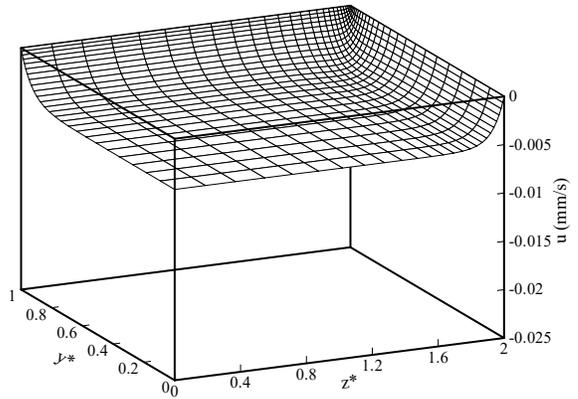

**Figure 2.** Velocity distribution for two different values of n and pH while keeping $\alpha = 2$ and $C_{KCl} = 0.005$.

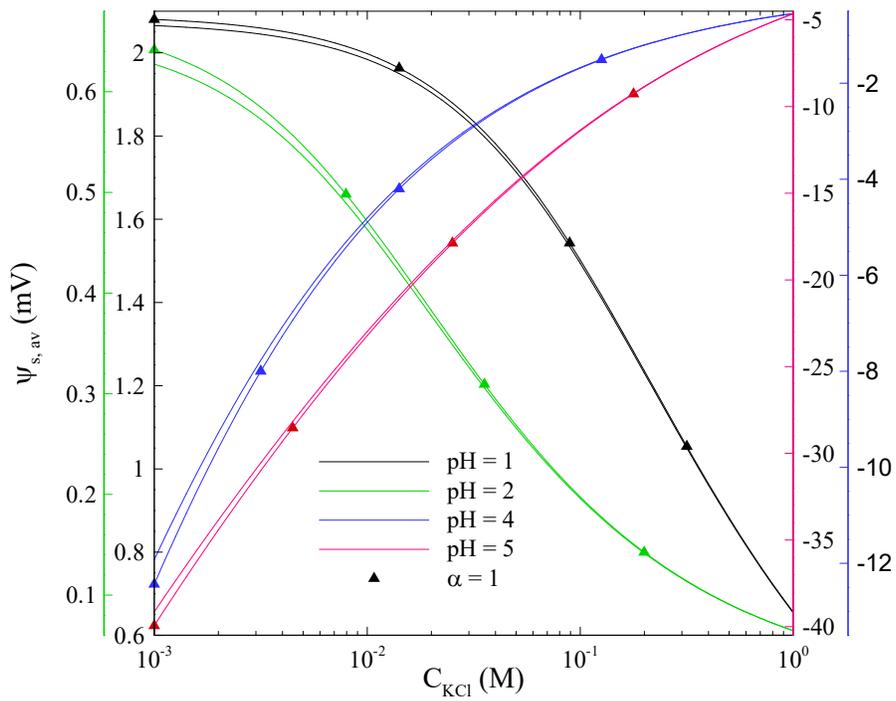

**Figure 3.** Variation of average surface potential versus background salt concentration at various values of pH and two different values of $\alpha = 1, 2$. Lines without symbols are related to $\alpha = 2$.

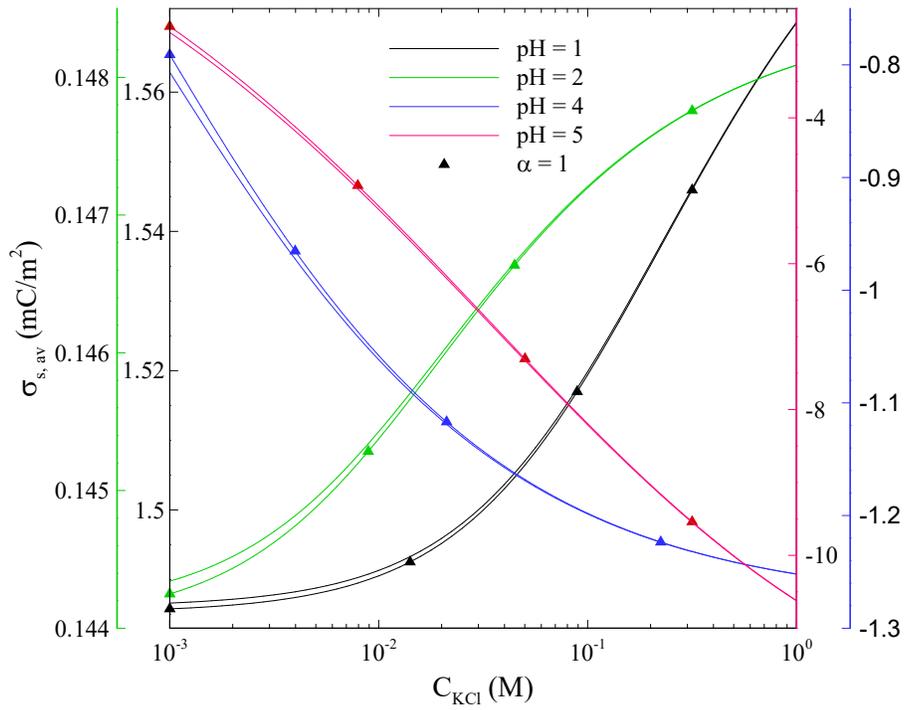

**Figure 4.** Variation of average surface charge density versus background salt concentration at various values of pH and two different values of $\alpha = 1,2$. Lines without symbols are related to $\alpha = 2$.

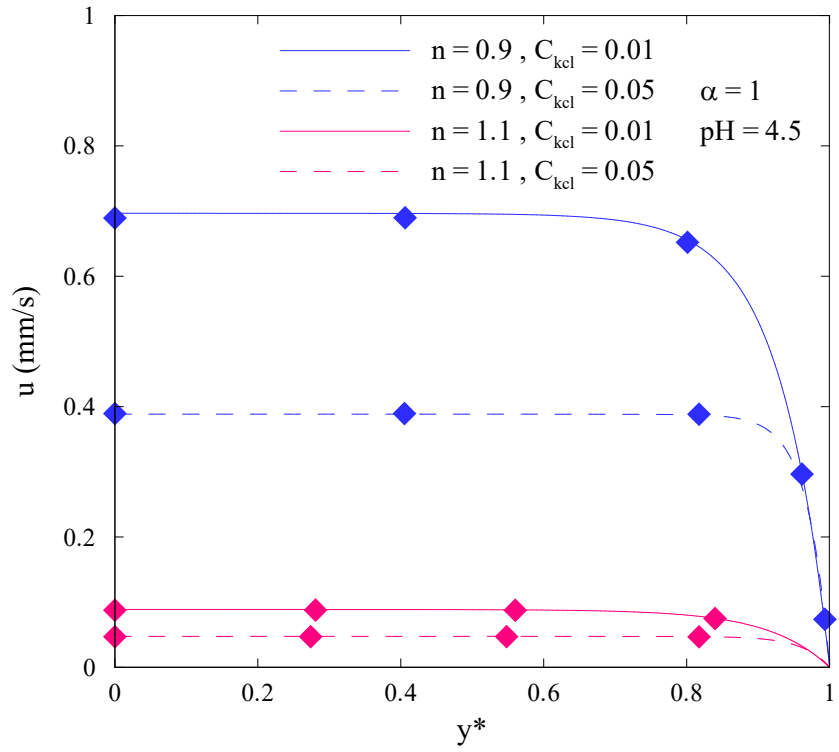

**Figure 5.** Velocity profile across the channel height at the channel center for two different values of n and $C_{KCl}$. Symbols are related to the results obtained from COMSOL Multiphysics.

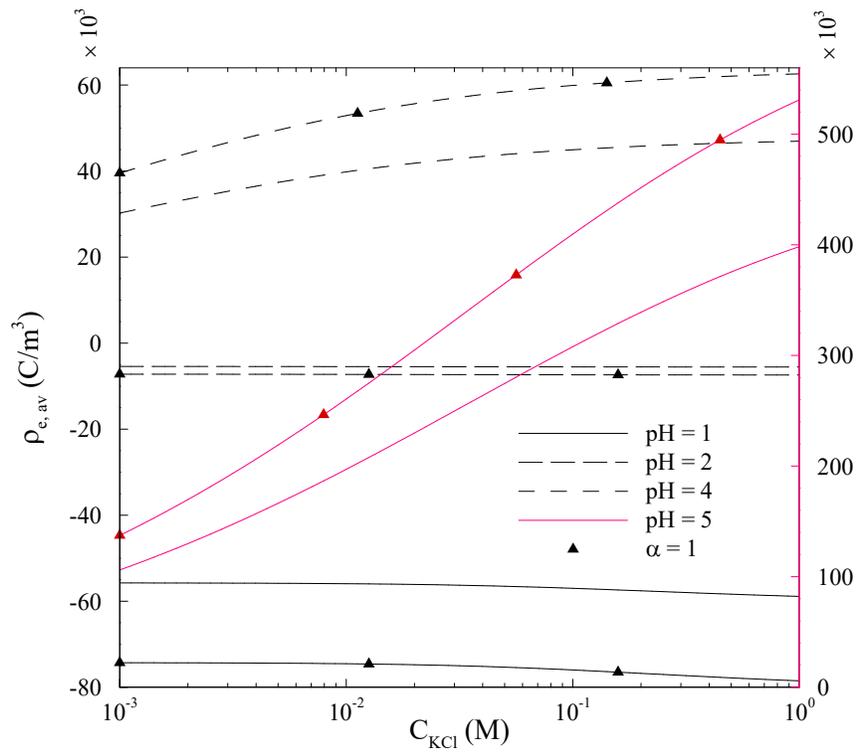

**Figure 6.** Effect of background salt concentration on average of electric charge at various values of pH and two different values of $\alpha = 1,2$. Lines without symbols are related to $\alpha = 2$.

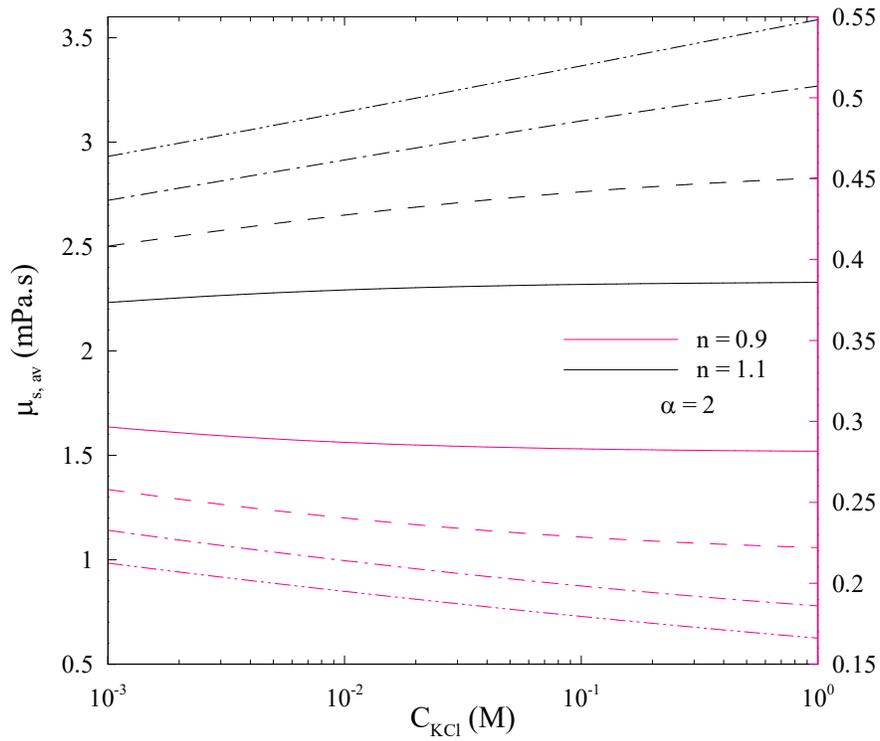

**Figure 7.** Average of viscosity at the channel surface versus background salt concentration at various values of pH and two different values of n = 0.9, 1.1. Solid, dashed, dash dot, and dash dot dot lines are related to pH = 4, 5, 6 and 7 respectively.

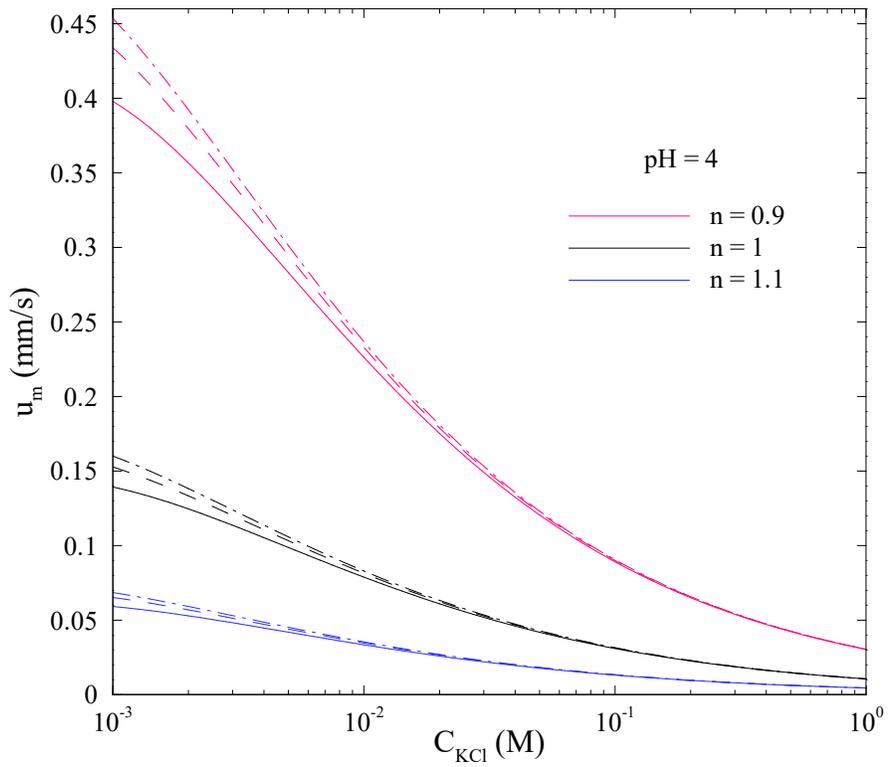

**Figure 8**. Mean velocity versus background salt concentration at various values of channel aspect ratio and flow behavior index. Solid, dashed and dash dot lines are related to $\alpha = 1$, 2 and 5 respectively.

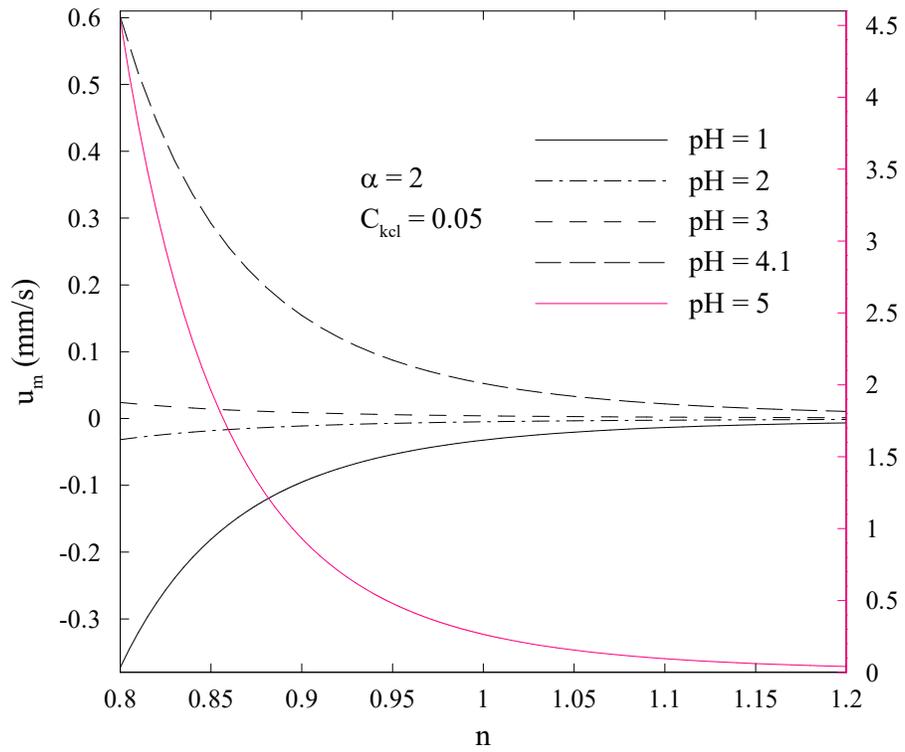

**Figure 9.** Variation of mean velocity with flow behavior index at various values of solution pH.

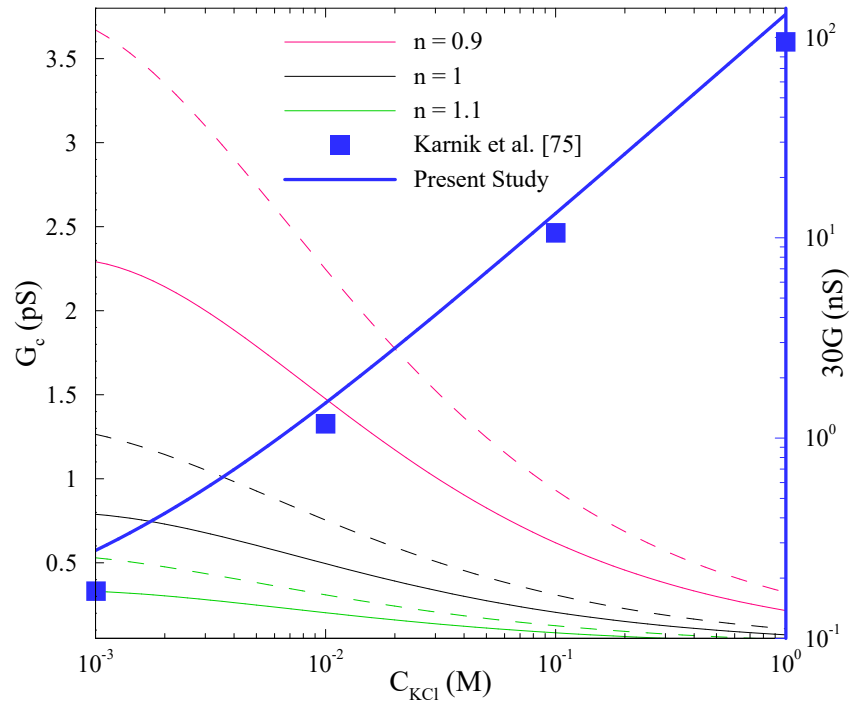

**Figure 10.** Influence of background salt concentration on convective ionic conductance at various values of flow behavior index while keeping pH=4. Solid and dashed lines are related to $\alpha =1$ and 2 respectively. The values of the ionic conductance multiplied by 30 for H=35nm, W=1μm, L=120μm and pH=5.5 are shown on the right part of the figure to be compared with the experimental data of Karnik et al. [75].

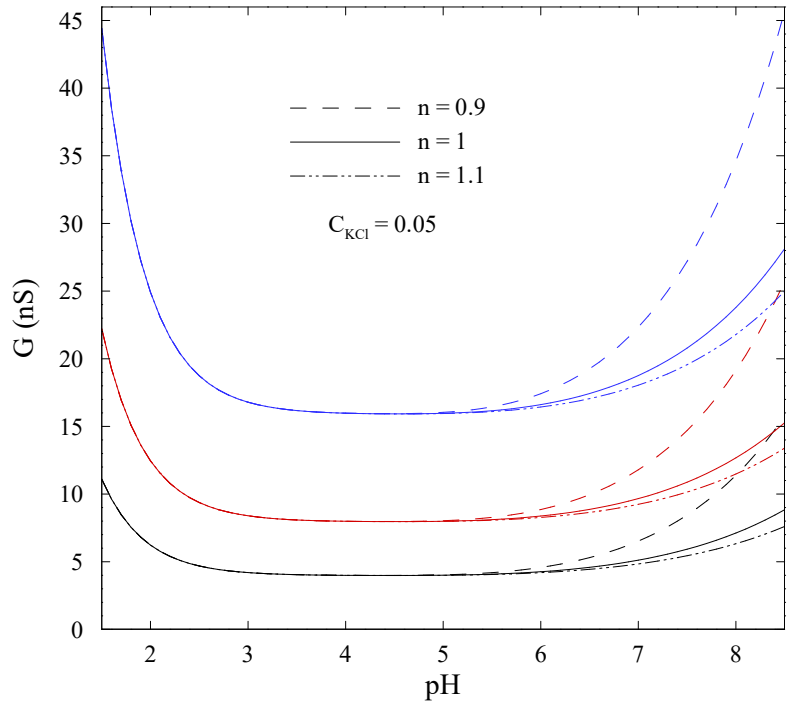

**Figure 11.** Ionic conductance versus solution pH at various values of flow behavior index and channel aspect ratio. Colors of black, red and blue are related to α =1, 2 and 4, respectively.

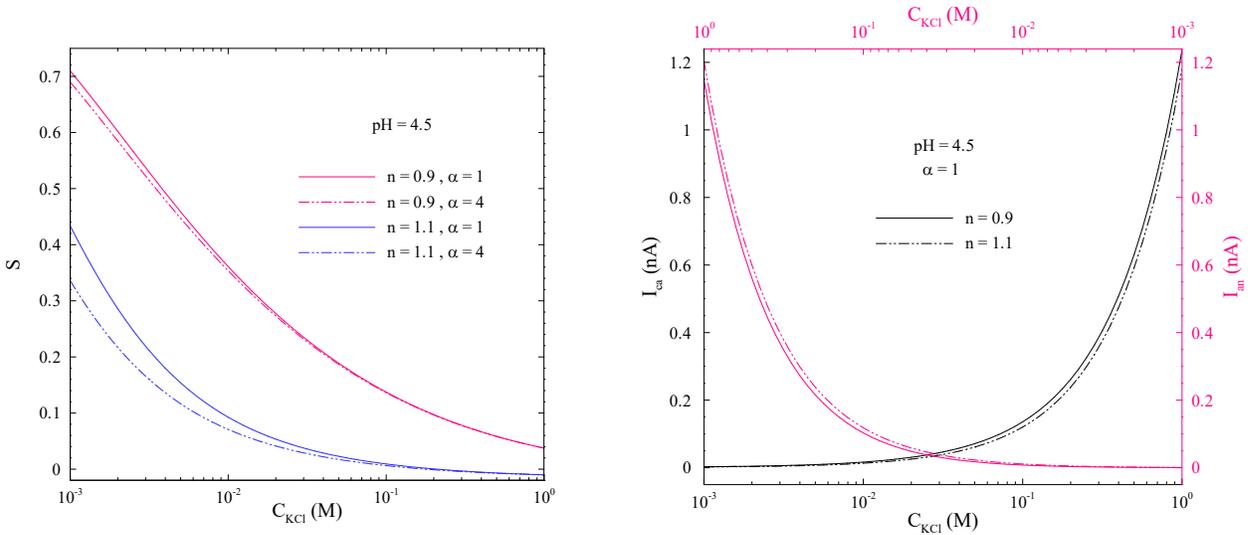

**Figure 12.** (a) Variation of ion selectivity versus background salt concentration at pH = 4.5 for four different combinations of $\alpha = 1, 4$ and $n = 0.9, 1.1$. (b) Ionic current versus background salt concentration at pH = 4.5 considering the same different combinations of part (a).

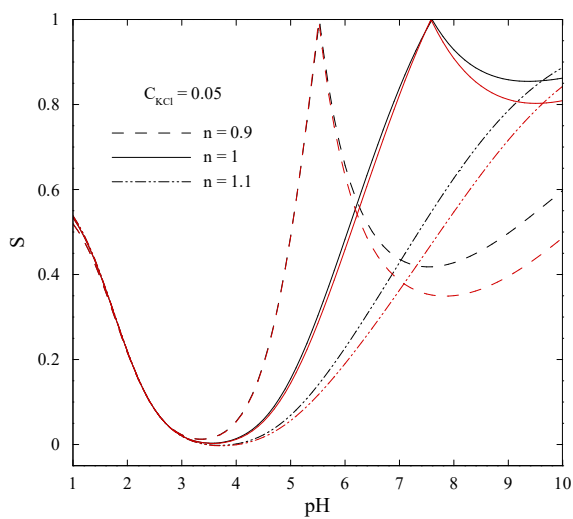 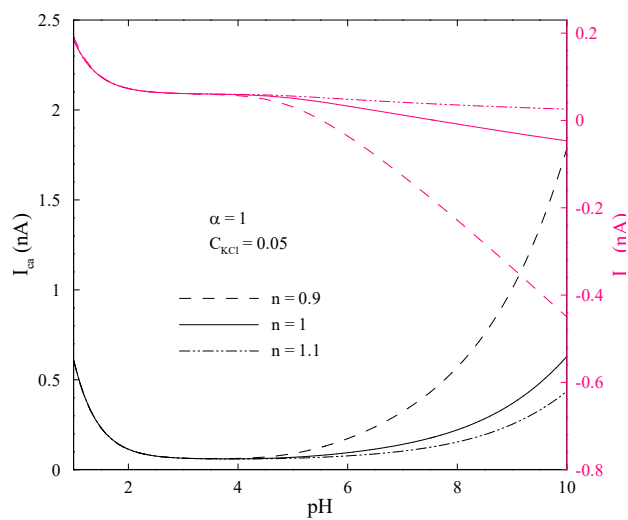

(a)                                    (b)

**Figure 13.** (a) Variation of ion selectivity versus solution pH for different values of flow behavior index. Colors of black and red are related to $\alpha = 1$, and 4 respectively. (b) Ionic current versus solution pH for different values of flow behavior index. The colors show the same aspect ratios in part (a).

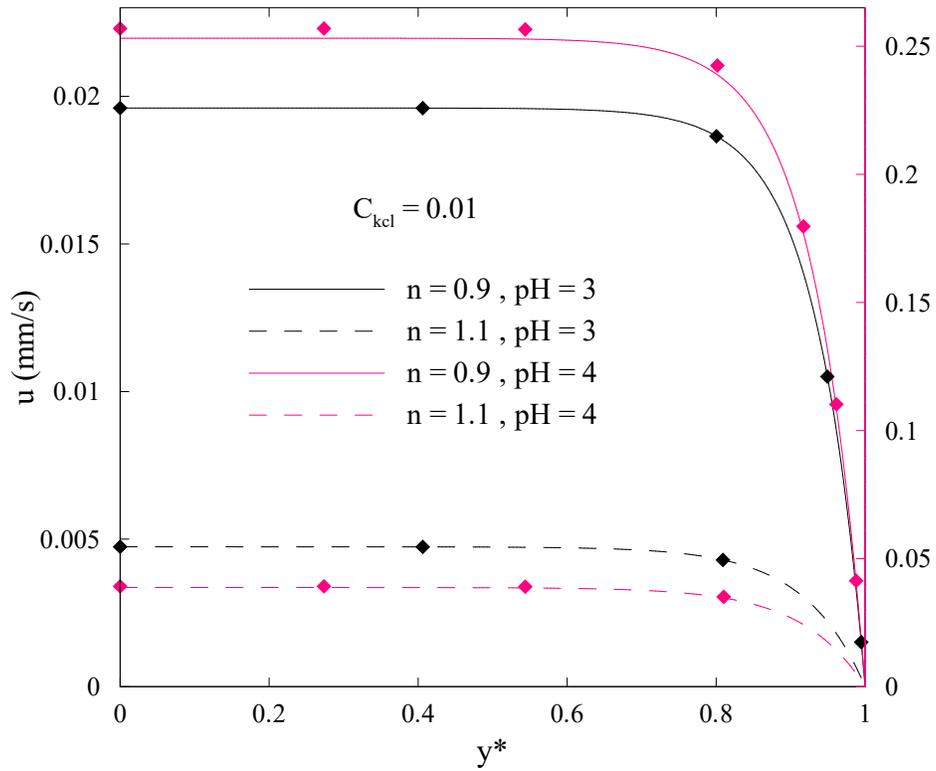

**Figure A1**. Comparison between the velocity profiles in a slit nanochannel obtained from analytical solution, Eq. (A14), against those calculated from numerical method of section 2.2 for a rectangular nanochannel with $\alpha = 10$. Symbols are related to the results of numerical method.